\documentclass[conference]{IEEEtran}
\pdfoutput=1
\let\labelindent\relax 

\usepackage{graphicx}
\usepackage{amsmath}
\usepackage{amsfonts}
\usepackage{amssymb}
\usepackage{xcolor}
\usepackage{enumitem}
\usepackage{url}
\usepackage{listings}
\usepackage{multirow}

\newcommand{\titletext}{CLTune: A Generic Auto-Tuner for OpenCL Kernels}

\definecolor{commentcolour}{HTML}{888888}
\begin{document}

\title{\titletext}

\author{\IEEEauthorblockN{Cedric Nugteren \qquad Valeriu Codreanu}
        \IEEEauthorblockA{SURFsara HPC centre, Amsterdam, The Netherlands \\
                         \url{mail@cedricnugteren.nl} \qquad \url{valeriu.codreanu@surfsara.nl}}
                        }

\maketitle
\begin{abstract}

This work presents CLTune, an auto-tuner for OpenCL kernels. It evaluates and tunes kernel performance of a generic, user-defined search space of possible parameter-value combinations. Example parameters include the OpenCL workgroup size, vector data-types, tile sizes, and loop unrolling factors. CLTune can be used in the following scenarios: 1) when there are too many tunable parameters to explore manually, 2) when performance portability across OpenCL devices is desired, or 3) when the optimal parameters change based on input argument values (e.g. matrix dimensions). The auto-tuner is generic, easy to use, open-source, and supports multiple search strategies including simulated annealing and particle swarm optimisation.
CLTune is evaluated on two GPU case-studies inspired by the recent successes in deep learning: 2D convolution and matrix-multiplication (GEMM). For 2D convolution, we demonstrate the need for auto-tuning by optimizing for different filter sizes, achieving performance on-par or better than the state-of-the-art. For matrix-multiplication, we use CLTune to explore a parameter space of more than two-hundred thousand configurations, we show the need for device-specific tuning, and outperform the clBLAS library on NVIDIA, AMD and Intel GPUs.


\end{abstract}



\section{Introduction}

The programmability of parallel processors such as graphics processing units (GPUs), multi-core CPUs and accelerators (MIC, FPGAs) has been improved significantly over the past years. Programming languages have matured, development tools have increased capabilities, libraries were created or improved, and so on. These new and improved technologies have led to highly efficient parallel programs in domains such as quantum chemistry, astronomy, bioinformatics, machine learning and fluid dynamics.

Despite these advancements, achieving close-to-peak performance remains a task for expert programmers in many cases~\cite{Du2012}. And even experts might face optimisation problems where the space of design decisions is too large to explore. Or they might have to tune their code for a wide variety of devices, each with their own sensitivities to specific parameters. Furthermore, to achieve optimal performance, code can be tailored to specific input arguments (e.g. matrix dimensions), either at compile-time (off-line) or at run-time (on-line).

We present CLTune, an auto-tuner designed to address the above issues for both experts and non-experts. We choose OpenCL as a target programming language, as it is supported by almost all recent accelerators (GPUs, MIC, FPGAs) and even by processors with less parallelism (x86 CPUs, ARM CPUs). Although OpenCL code is portable across devices, it is definitely not performance portable: to achieve good performance it is necessary to tune design parameters, adjust the hierarchy of parallelism, and explore different algorithms~\cite{Du2012}. With CLTune, users can specify tunable parameters along with a list of values for their OpenCL kernels. Our auto-tuner is designed for the following scenarios:

\subsubsection{The search-space is too large to explore manually} For example, consider a case where the vector width for the input and output data can be varied, a 2D rectangular local \textit{workgroup} is used (\textit{threadblock} in CUDA terminology), the thread-coarsening factor (i.e. the amount of work per thread) can be adjusted, and a decision can be made whether or not to cache in \textit{local} memory (\textit{shared} memory in CUDA terminology). In such a case, the number of permutations will quickly grow into the thousands. Exploring this search-space is especially important for library designers: they are willing to spend extra effort to achieve the best possible performance.

\subsubsection{Efficient execution on a variety of devices is desired} For performance portability across devices, data-locality and algorithmic choices might impact performance the most. However, beyond that, even for devices coming from the same vendor or with the same architectural family, it might be worthwhile to explore workgroup configurations, loop unroll factors, or vector widths~\cite{Du2012}.

\subsubsection{The optimal parameters change based on input arguments} Example arguments are data-sizes or boolean variables. A use-case is an OpenCL kernel executed in a time-loop as part of a scientific simulation with a run-time fixed input size: perhaps the first tens of time-steps can be used to find optimal parameters, allowing the remainder time-steps to execute more efficiently. Another use-case is a fast Fourier transform (FFT) library of which the parameters can be tuned at compile-time when it is known which FFT-sizes are being used (e.g.~\cite{Li2013}).

The contributions of this work are two-fold. First, we present CLTune, a new OpenCL-kernel auto-tuner which automatically explores a search-space of tuning parameters (section~\ref{tuner}). The tuner is designed to be generic, easy to use, flexible and customisable. CLTune is furthermore open-source and supports different search strategies: full-search, random-search, simulated annealing, and particle swarm optimisation. Second, we present two case-studies inspired by the recent successes with convolutional neural networks for deep-learning on GPUs~\cite{Jia2014,Krizhevsky2012}: 2D convolution and matrix-multiplication (GEMM). For 2D convolution, our tuner matches the CUDA-based state-of-the-art~\cite{Werkhoven2014} and achieves up to 1658 GFLOPS on a 11x11 filter and 207GB/s on a 3x3 filter running on an AMD HD7970 GPU. We show that tuning for specific filter sizes is worthwhile, yielding a performance boost of up to 56\%. We also tune matrix-multiplication, matching the state-of-the-art~\cite{Matsumoto2014}, and achieving better performance compared to the clBLAS library. Thanks to CLTune, our GEMM kernel is the best-performing publicly available OpenCL implementation. We furthermore demonstrate that tuning for each device individually can lead to a factor 2 performance increase.

\section{Related work}

There are many existing auto-tuners, of which ATLAS might be the best-known example. In the first category of related work, we discuss cases where auto-tuners were introduced to solve a specific OpenCL or GPU-related problem. Examples are auto-tuners for convolution~\cite{Werkhoven2014}, sparse matrix-vector multiplications~\cite{Grewe2011}, dense matrix-matrix multiplications~\cite{Li2009,Matsumoto2012}, and FFTs~\cite{Li2013}. In contrast to CLTune, such tuners are tailored for a specific problem and are often not applicable beyond these problems. A more generic OpenCL auto-tuner is Maestro data-orchestration tuner~\cite{Spafford2010}, however it is orthogonal to this work since it works on data transfers rather than kernel. We are not aware of any existing \textit{generic} tuner for OpenCL kernels.

There are however existing tuners which do target GPUs but work on a higher-level representation than OpenCL. Examples include tuners for a GPU skeleton programming framework~\cite{Enmyren2010a}, as part of a parallelizing compiler~\cite{Khan2013}, for OpenMP 4.0 directives~\cite{Lee2010}, for HMPP directives~\cite{GrauerGray2012}, and for mathematical expressions in Theano~\cite{Bergstra2011}. Such tuners typically tune higher-level concepts and are thus not designed for fine-grained parameter tuning. Furthermore, such tuners are not as generic as CLTune: users cannot define and tune their own (low-level) parameters.

\section{CLTune: an auto-tuner for OpenCL kernels}
\label{tuner}

CLTune is an auto-tuner written in C++11 targeting OpenCL kernels. It is open-source and freely-available under the APACHE 2.0 license on GitHub\footnote{CLTune on GitHub: \url{https://github.com/CNugteren/CLTune}}, which includes the two case-studies of this work as example codes. The tuner is implemented as a library with a C++ API, and can thus be used as stand-alone for off-line (compile-time) tuning or integrated into existing source-code for on-line (run-time) tuning. The API is designed such that all host OpenCL API calls are hidden from the user, i.e. CLTune takes care of tasks such as OpenCL device initialisation, kernel invocation, and device memory management.

To illustrate the ease of use of the tuner, we briefly discuss a simple example. Consider the example OpenCL kernel at the top of Fig.~\ref{fig:code_example}, which uses parameter $WPT$ to vary the amount of work per thread. With this OpenCL kernel saved as the file \texttt{copy.cl}, CLTune can be used to explore performance for 1, 2 or 4 copies per thread. In the remainder of Fig.~\ref{fig:code_example}, a tuner is created and the copy-kernel added. The example uses the function \texttt{DivGlobalSize} to reduce the total number of threads as we increase the workload per thread. The tuner will compile three different versions of the copy-kernel (with $WPT$=\{1,2,4\}) and will invoke each with a fresh device-copy of the supplied host-arguments (\texttt{in\_vector} and \texttt{out\_vector}). The kernel execution will be timed and the the value of best $WPT$ value will be reported.

\begin{figure}[!t]
\begin{lstlisting}
  __kernel void copy(__global float* in,
                     __global float* out) {
    const int tid = get_global_id(0);
    for (int w = 0; w < WPT; ++w) {
      out[tid*WPT + w] = in[tid*WPT + w];
    }
  }
\end{lstlisting}
\begin{lstlisting}
  // Creates a new tuner on device 1 of platform 0
  cltune::Tuner tuner(0, 1);

  // Kernel: 2048/WPT global and 64 local threads
  tuner.AddKernel("copy.cl", "copy", {2048}, {64});
  tuner.AddParameter("WPT", {1, 2, 4});
  tuner.DivGlobalSize({"WPT"});

  // Specifies the input and output host arrays
  tuner.AddArgumentInput(in_vector);
  tuner.AddArgumentOutput(out_vector);

  // Starts the tuning process
  tuner.Tune();
\end{lstlisting}
\caption{Basic example of using CLTune on a kernel to copy 2048 elements.}
\label{fig:code_example}
\end{figure}

\subsection{Advanced usage}

As we have seen, using CLTune can be quite simple. Nonetheless, the tuner is also equipped with more advanced features, which are introduced in this section: 1) manipulation of the thread and workgroup configurations, 2) constraints on the parameters, and 3) result verification.

 The copy-example of Fig.~\ref{fig:code_example} could be extended by additionally tuning the local workgroup size. This is as easy as introducing a new parameter $WG$ with values of for example \{32, 64, 128, 256\}, changing the old workgroup size from 64 into 1, and using \texttt{MulLocalSize(\{"WG"\})} to multiply 1 by $WG$. The reason that the thread dimensions in the example code are within braces is the following: the example uses a 1D arrangement, but they can be extended to 2D or 3D.

Users of the auto-tuner might want to set constraints on parameter combinations. An example is tuning a 2D workgroup size $X_{wg}$ by $Y_{wg}$: a 128x4 or 4x128 configuration might be allowed, but 128x128 might not. In other words, $X_{wg} \cdot Y_{wg}$ has to remain below a certain threshold. Such constraints can be specified by the user (e.g. as lambda expressions) and can thus be as complex as needed. The two case-studies discussed in this paper make heavy use of such constraints. CLTune furthermore automatically imposes constraints based on the device limits (workgroup dimensions, local memory size, etc.).

CLTune also provides an interface to verify the results of each tested configuration. This can be used to make sure that all tested parameter permutations are indeed correct and no parameter-dependent bugs are present in the kernel. Verification only requires a reference kernel to be passed to the \texttt{SetReference} function in a similar way as \texttt{AddKernel} in Fig.~\ref{fig:code_example}. The outputs of each tested kernel are then automatically compared against the outputs of the reference kernel.

\subsection{Search strategies and search-space properties}

In some cases it is not feasible or desirable to explore all points in the search-space, i.e. all permutations of all possible parameter values. This is the case when the search-space is too large or when the tuning is done on-line, as in this case the tuning-time affects performance of the actual application. Note that tuning-time is not only determined by execution time of the OpenCL kernel, but can also be limited by the repeated re-compilation of the slightly modified OpenCL kernel. Three alternatives to the default full-search are available for those scenarios: random-search, simulated annealing, and particle swarm optimisation.

Random-search is the simplest of the search strategies: it samples and tests a random configurable fraction of the entire search-space and reports the best-found configuration. Its efficacy is dependent on the shape of the search-space: when the percentage of good parameter combinations is low, the chance to find one is low as well (and vice-versa).

Before discussing the other two search strategies, we discuss properties of the search-space and alternative search strategies. In theory, we cannot make any assumptions about the search-space, as it is defined completely by the user of the tuner. Still, there are some useful observations possible given the simple fact that CLTune tunes OpenCL kernels:
\begin{enumerate}
    \item The number of values a parameter can take is typically low, as shown in earlier examples and the two case-studies. Examples are vector widths (2 to 5), thread-coarsening factors (2 to 8), and workgroup sizes (up to 5 per dimension). Furthermore, some parameters might be boolean and have only 2 possible values such as: unroll loops or not, use local memory or not.
    \item The search-space can be highly dimensional. A 3 or 4 dimensional space is easily created in a simple example, and even a 14 dimensional one is possible as illustrated by the matrix-multiplication case-study.
    \item Parameters are typically discrete and have many non-linearities. For example, performance might increase when increasing the work per thread from 1 to 2 and from 2 to 4, but might decrease dramatically from 4 to 8 as register pressure suddenly becomes a bottleneck.
    \item There can be a strong relation between parameters, e.g. a small number of threads in one workgroup dimension might only be useful if the number of threads is large in the second dimension.
\end{enumerate}

Because of the expected non-linearities and boolean variables, methods based on derivatives or automatic differentiation are not applicable. Even \textit{derivative free} methods such as direct search~\cite{Kolda2003} are not suitable, since they assume that it is relatively cheap to explore all neighbours of a particular configuration. If the search-space is indeed narrow and highly dimensional, this is certainly not the case, as many or all configurations are (diagonal) neighbours of each other. Therefore, taking into account the above 4 observations, we focused on heuristics and iterative methods such as simulated annealing and particle swarm optimisation first. However, other search methods are easily pluggable into CLTune, so evolutionary search, gradient methods, stochastic optimisation or dynamic programming can be evaluated as part of future work.

\subsection{Simulated annealing}

Simulated annealing (SA) is a heuristic search method inspired by annealing in metallurgy~\cite{Kirkpatrick1983} which iteratively moves through the search-space from neighbour to neighbour and ends after a fixed number of iterations or when a certain criterion is reached. In principle, simulated annealing only moves to neighbours with better performance (\textit{lower energy} in annealing terminology). However, to prevent getting stuck in a local optimum, the heuristic has a certain probability to make a step towards a worse configuration. This probability decreases over time as the annealing \textit{temperature} decreases: at the end of the search the likelihood to move towards the global optimum is higher than at the start. This probability is further decreased as a function of the difference in performance between the current and a neighbour configuration.

The simulated annealing heuristic is implemented in CLTune. The search is initialized in a random configuration and continues until a user-defined number of configurations have been explored. At each step, a random neighbour $s'$ of the current configuration $s$ is chosen and its performance is evaluated. This neighbour has a probability $P$ of becoming the new current state:
\begin{equation*}
  P(t,t',T)=\begin{cases}
    1 & \text{if $t'<t$}\\
    e^{ -(t' - t)\cdot T^{-1} } & \text{otherwise}
  \end{cases}
\end{equation*}
in which $T$ is the annealing temperature, and $t$ and $t'$ represent the execution times of configurations $s$ and $s'$ respectively. The performance of simulated annealing in CLTune and its sensitivity to the parameter $T$ are evaluated for the case-studies in sections~\ref{sec:conv} and~\ref{sec:gemm}.

\subsection{Particle swarm optimisation}

Particle swarm optimisation (PSO) is an evolutionary search strategy in which a \textit{swarm} of $S$ communicating \textit{particles} explore the search-space~\cite{Kennedy1995}. These particles are initially positioned randomly and are given a random velocity. At each step $t$, the new positions $x_{i}^{t+1}$ of each particle $i$ are calculated based on their current positions $x_{i}^{t}$ and their velocities $v_{i}^{t}$. These velocities are in turn updated as a function of their current velocities, their own best-known configuration so-far $p_{i}^{t}$, and the overall best-known configuration so far $g^{t}$. A variant of PSO is accelerated PSO~\cite{Yang2011}, in which the new position in each dimension $d$ is directly calculated based on the old position, removing the concept of velocity:
\begin{equation*}
    x_{i,d}^{t+1} = \alpha\epsilon_{d} + \beta p_{i,d}^{t} + \gamma g_{d}^{t} + (1-\alpha-\beta-\gamma)x_{i,d}^{t}
\end{equation*}
in which $\alpha$, $\beta$ and $\gamma$ are probability parameters and $\epsilon_{d}$ represents a random number within the range of the parameter in dimension $d$.

CLTune uses a modified version of accelerated PSO. Since the search-space might be narrow and discrete, it makes little sense to compute a linear combination of fractions of other positions and round it to an integer. Instead, we omit the notion of continuity and compute the new position $x_{i,d}^{t+1}$ in each dimension $d$ as follows:
\begin{equation*}
x_{i,d}^{t+1}=\begin{cases}
    \epsilon_{d} & \text{with probability $\alpha$ (random)}\\
    p_{i,d}^{t} & \text{with probability $\beta$ (local best)}\\
    g_{d}^{t} & \text{with probability $\gamma$ (global best)}\\
    x_{i,d}^{t} & \text{otherwise (current position)}
  \end{cases}
\end{equation*}
with $\alpha+\beta+\gamma \le 1$. Note that this is applied for each dimension separately: the likelihood of moving to a completely unrelated position is for example only $\alpha$ to the power of $d$. The performance of PSO is evaluated in sections~\ref{sec:conv} and~\ref{sec:gemm}.

\section{Experimental setup}

In sections~\ref{sec:conv} and \ref{sec:gemm} we will perform experiments on OpenCL GPUs from different vendors, chosen for their wide architectural variety. These devices and their relevant properties are listed in table~\ref{tbl:setup}. We plan to extend our experiments in future work to non-GPU OpenCL devices (CPUs, MIC).

All data-types in this paper are single-precision floating point numbers. Furthermore, simulated annealing is configured with a variable temperature $T=\{2,4,6\}$, and PSO is configured with $\alpha=0.4$, $\beta=0$ (no local-best influence as argued by~\cite{Yang2011}), $\gamma=0.4$ and a variable swarm size $S=\{3,6\}$.

\begin{table}[!ht]
  \setlength{\tabcolsep}{4pt}
  \centering
  \caption{Experimental setup of the OpenCL devices}
  \begin{tabular}{lll|ccc|}
    vendor and            & archi-    & compiler     & peak        & peak      & GFLOPS   \\
    device name           & tecture   & and SDK      & GFLOPS      & GB/s      & per GB/s \\
    \hline
    NVIDIA Tesla K40m     & Kepler    & CUDA 7.0     & 4291        & 288       & 14.9 \\
    NVIDIA GeForce GTX480 & Fermi     & CUDA 5.5     & 1345        & 177       & 7.6  \\
    AMD Radeon HD7970     & Tahiti    & APP 2.9      & 4368        & 288       & 15.1 \\
    Intel Iris 5100       & Iris      & Apple 2.4.2  &  832        &  26       & 32.5 \\
  \end{tabular}
  \label{tbl:setup}
\end{table}

\section{Case-study: 2D convolution}
\label{sec:conv}

Our first case-study is 2D convolution, which is for example for image filters such as blur or edge detection. This case-study is motivated on one hand by the conceptual simplicity of convolution, and on the other hand by the recent renewed interest in GPU-based convolution within the context of convolutional neural networks for deep-learning. Convolution is used in production as well as research tools and libraries, such as Theano~\cite{Bergstra2011}, Caffe~\cite{Jia2014}, cuDNN, Torch7, and cuda-convnet~\cite{Krizhevsky2012}. In convolutional neural networks, many 2D convolutions have to be performed, typically with a range of compile-time fixed filter sizes, ranging for example from 3 by 3 to 11 by 11. An auto-tuner can be used to optimize the convolution code for each specific filter size, as is done for example by Theano. Auto-tuning will become even more important when moving from CUDA to OpenCL to enable execution on a wider range of devices. Convolution is also important outside the context of deep-learning, as is discussed in~\cite{Qadeer2013}, in which it is also shown that convolution can be generalized to other image and video processing operations such as SAD and SIFT.

\begin{figure}[!ht]
    \centering
    \includegraphics[width=0.82\columnwidth]{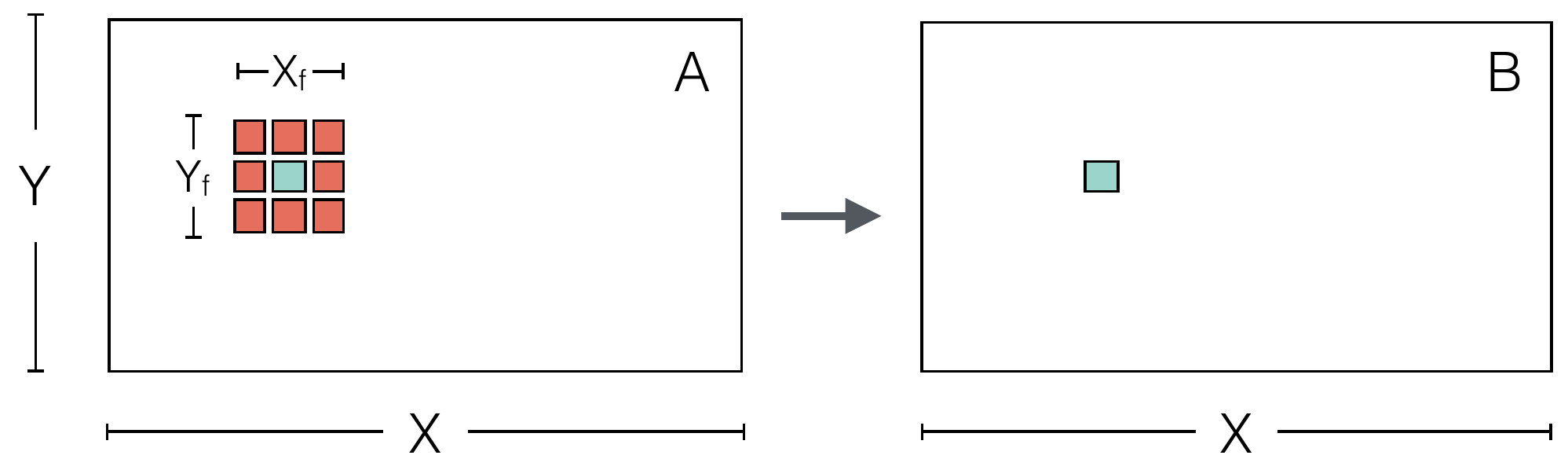}
    \caption{Illustration of an $X_{f}$ by $Y_{f}$ 2D convolution on an $X$ by $Y$ image.}
    \label{fig:conv1}
\end{figure}

The 2D convolution operation is illustrated in Fig.~\ref{fig:conv1}. In this figure, an $X_{f}$ by $Y_{f}$ filter $F$ is applied onto an $X$ by $Y$ input image $A$, producing an output image $B$ of the same size. For each $x,y$ coordinate, the output value can be calculated as follows in case $X_{f}=Y_{f}=3$:
\begin{equation*}
    B_{x,y} = w\cdot\sum\limits_{i=-1}^{i\le 1}\sum\limits_{j=-1}^{j\le 1}F_{i,j}A_{x+i,y+j}
\end{equation*}
in which $w$ is a weighing factor. The user-defined filter coefficients $F$ are the same for each coordinate and are therefore a good match for OpenCL's \textit{constant} memory. 

\subsection{Tuning parameters}

\begin{figure}[!t]
    \centering
    \includegraphics[width=0.59\columnwidth]{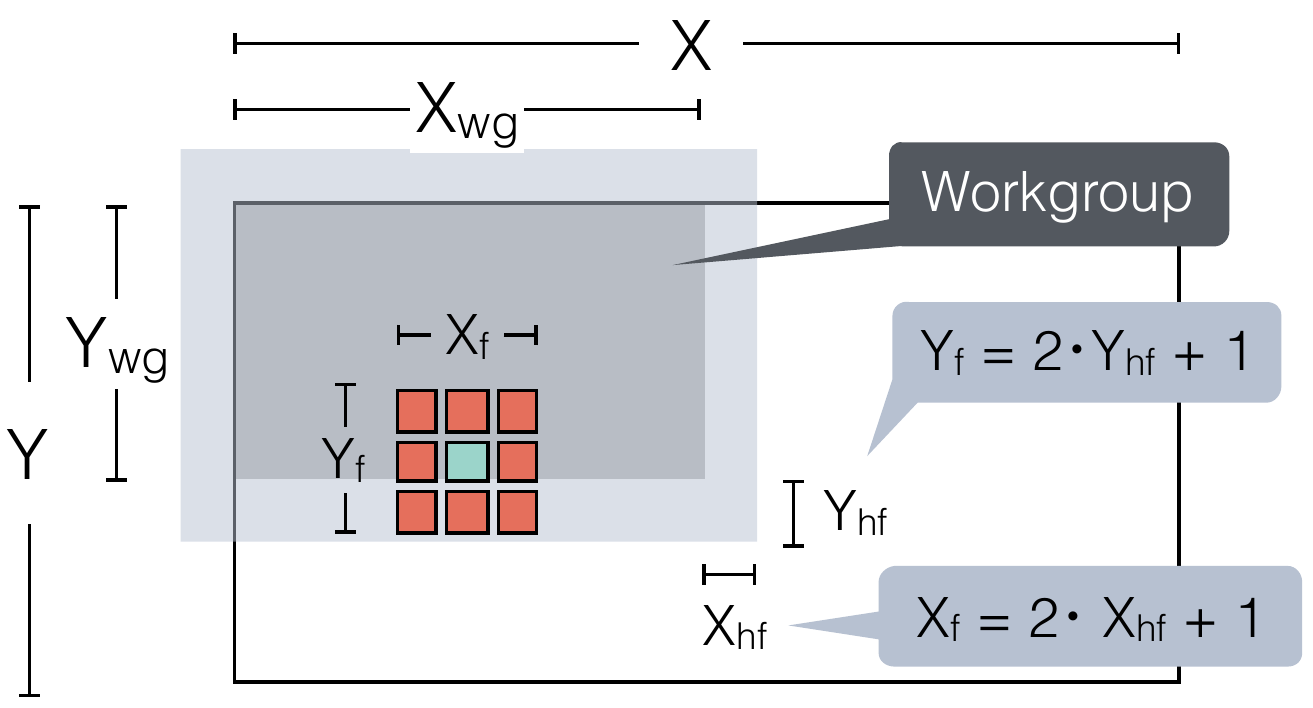}
    \includegraphics[width=0.39\columnwidth]{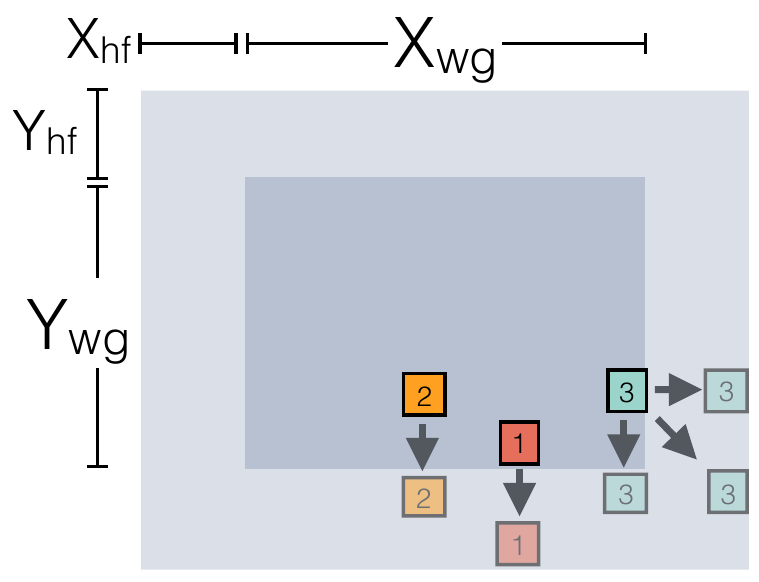}
    \caption{Division of the 2D convolution workload onto workgroups (left) and caching into local memory with $L\$ = 1$ and $X_{hf}=Y_{hf}=2$ (right).}
    \label{fig:conv2}
    \label{fig:conv3}
\end{figure}

We implemented a highly tunable implementation of 2D convolution in OpenCL inspired by~\cite{Podlozhnyuk2007a} and~\cite{Werkhoven2014}. Details of our implementation are not the focus of this work, but the source is available for inspection as part of the CLTune examples and further discussion can be found in related work~\cite{Werkhoven2014}. We do discuss the chosen tuning parameters:
\begin{itemize}
    \item The total amount of work is divided into rectangular local workgroups of size $X_{wg}$ by $Y_{wg}$, each tunable individually. This is illustrated by Fig.~\ref{fig:conv2}.
    \item The amount of work per thread $X_{wpt}$ and $Y_{wpt}$ is configurable in both dimensions. This results in a per-thread rectangular block of loads, computations and stores. Advantages might come from improved data-locality and re-use of filter coefficients and indexing variables.
    \item The parameter $L\$$ implements 3 caching strategies:
    \begin{enumerate}[start=0]
        \item Caching is hardware-only, no local memory.
        \item Input data is cached into the local memory as follows: every thread caches the value at its coordinates $x,y$ plus optionally up to three values from the halo of size $X_{hf}$ and $Y_{hf}$. This is illustrated in Fig.~\ref{fig:conv3}, in which the red example-thread (1) and blue example-thread (2) each load one additional halo value. The orange example-thread (3) loads 3 values since it is within $X_{hf}$ and $Y_{hf}$ of the $x$ and $y$ borders respectively.
        \item Extra helper threads are launched, creating workgroups of $X_{wg}+2\cdot X_{hf}$ by $Y_{wg}+2\cdot Y_{hf}$. Now, each thread caches only a single value (or more depending on the work per thread parameters). After caching into the local memory, the extra halo threads are canceled: they don't perform any computations.
    \end{enumerate}
    \item If $X_{wpt}$ is larger or equal than the vector width $VW$, multiple stores in the x-dimension can be combined into a single vector operation. Vector loads are only applied for the second local-memory strategy ($L\$=2$).
    \item The local memory is padded by $PAD$ in in the x-dimension to possibly avoid memory bank-conflicts.
    \item Unrolling of the $i$ and $j$ loops over the filter coefficient is enabled or disabled by $UNR$.
\end{itemize}

\subsection{Evaluation of the search strategies}

\begin{figure*}[!t]
    \centering
    \includegraphics[width=0.32\textwidth]{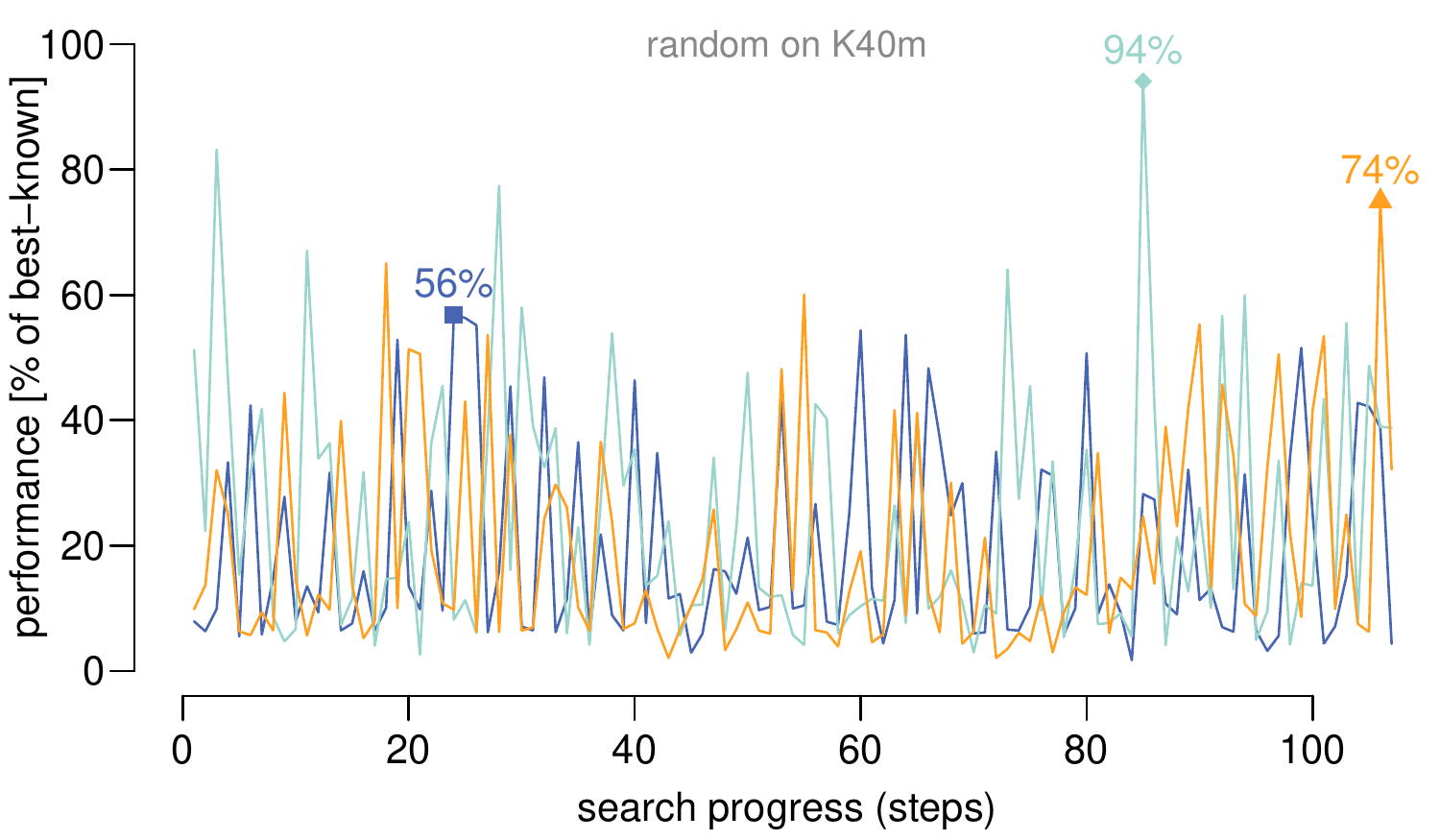}
    \includegraphics[width=0.32\textwidth]{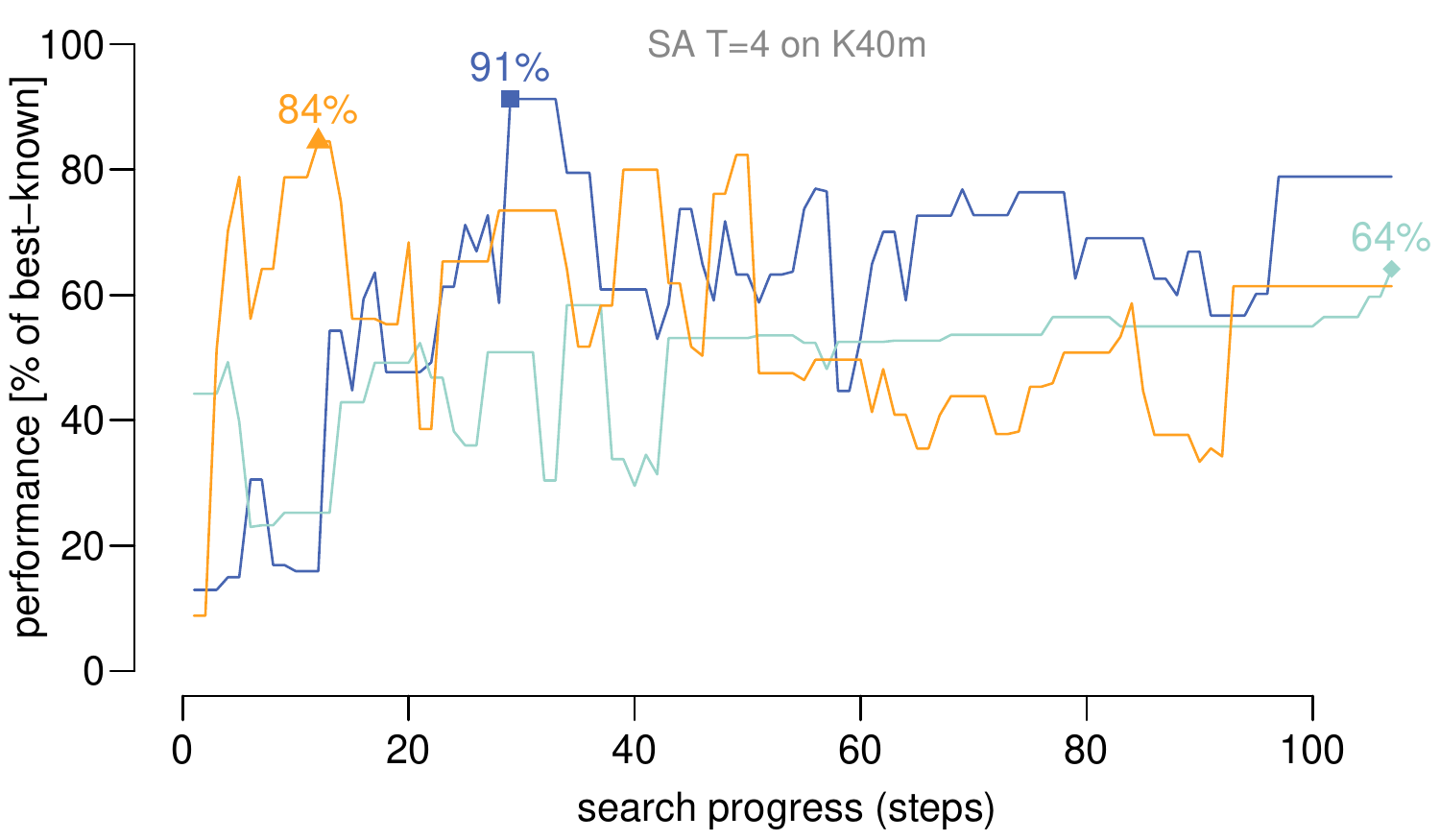}
    \includegraphics[width=0.32\textwidth]{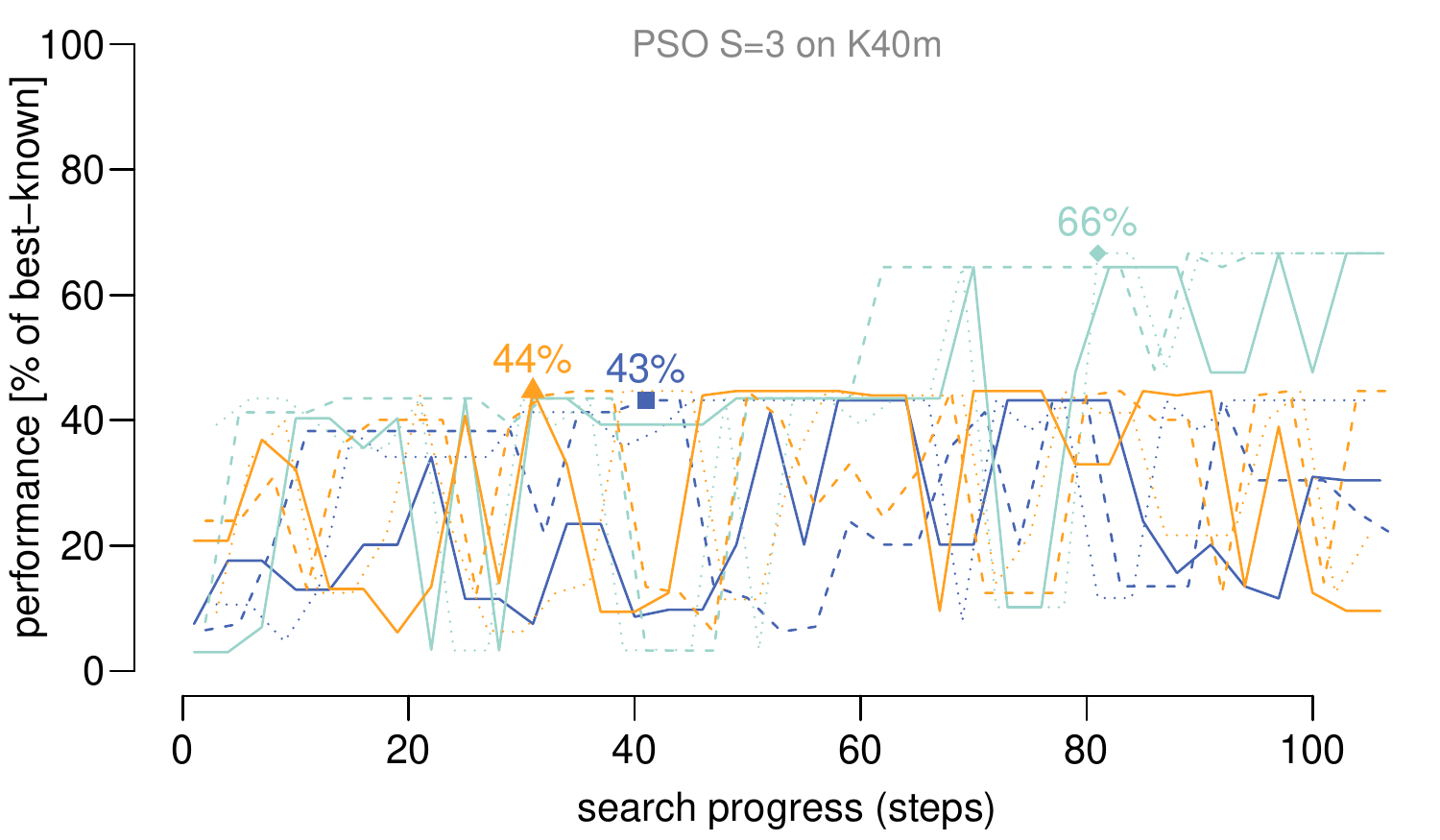}
    \caption{Search progress of search strategies on the Tesla K40m: random-search (left), simulated annealing (middle) and PSO (right). Because of the randomness involved in all strategies, we show multiple runs in different colours. For PSO, we distinguish the 3 different particles by solid, dashed and dotted lines.}
    \label{fig:conv_trace}
\end{figure*}

\begin{figure*}[!t]
    \centering
    \includegraphics[width=0.48\textwidth]{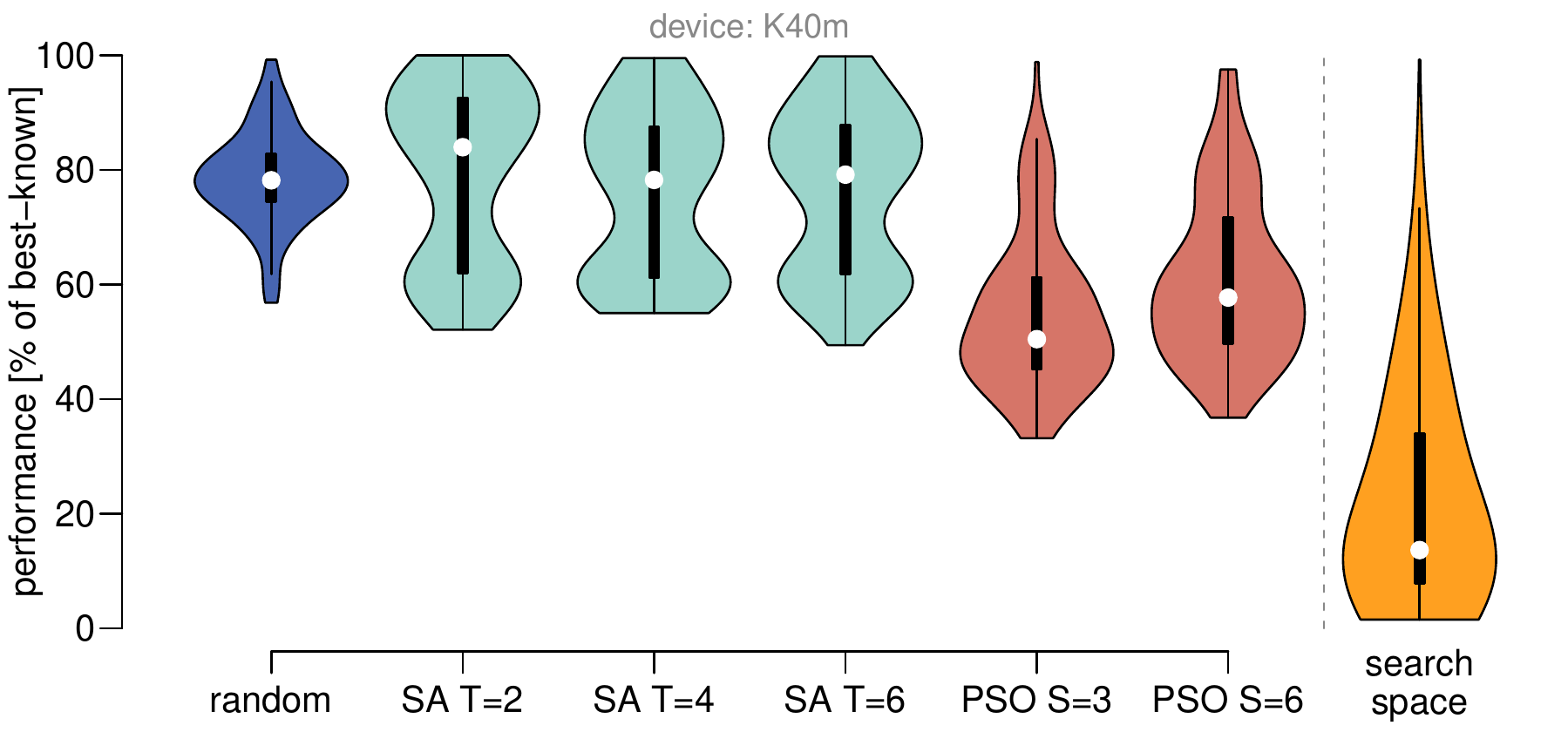}
    \includegraphics[width=0.48\textwidth]{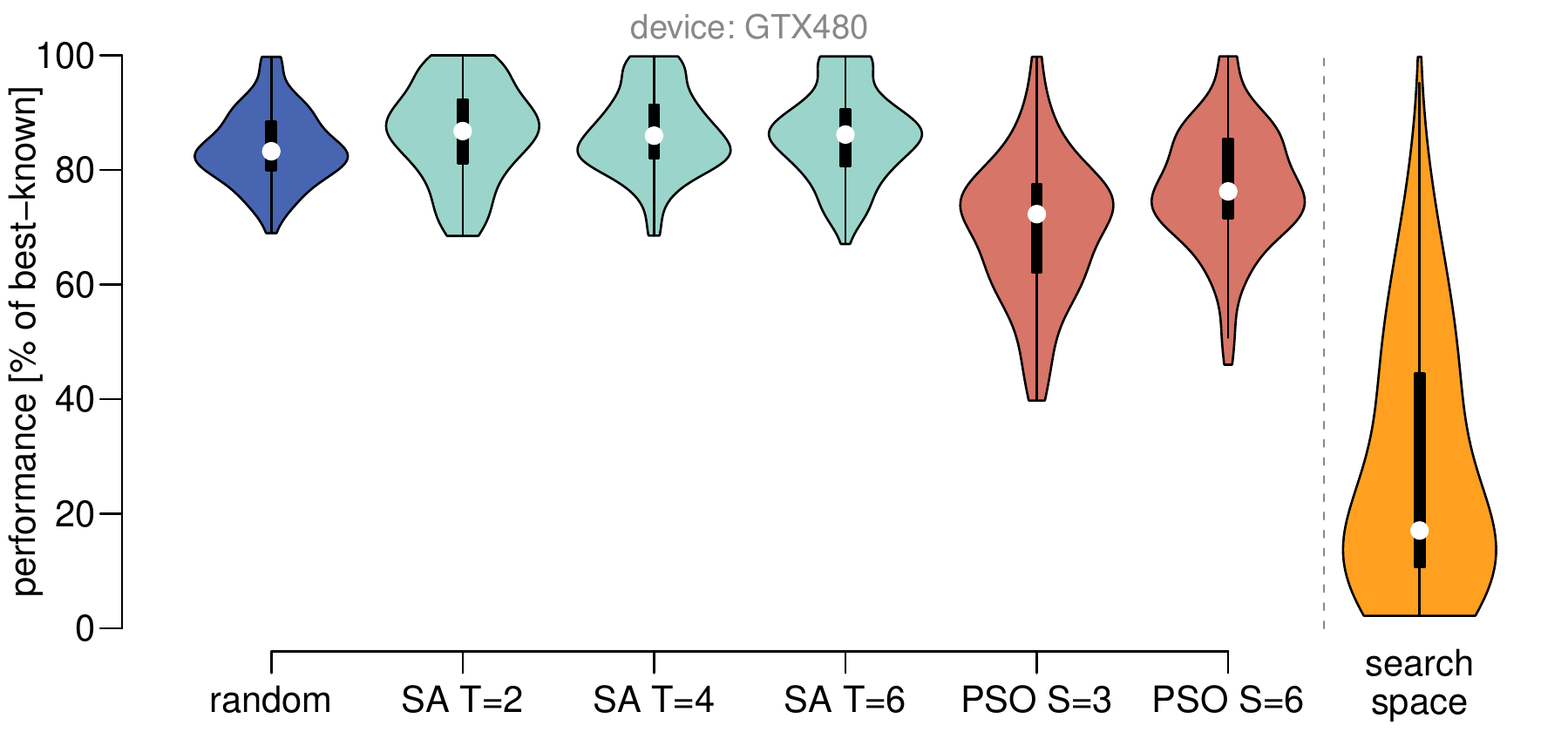}
    \includegraphics[width=0.48\textwidth]{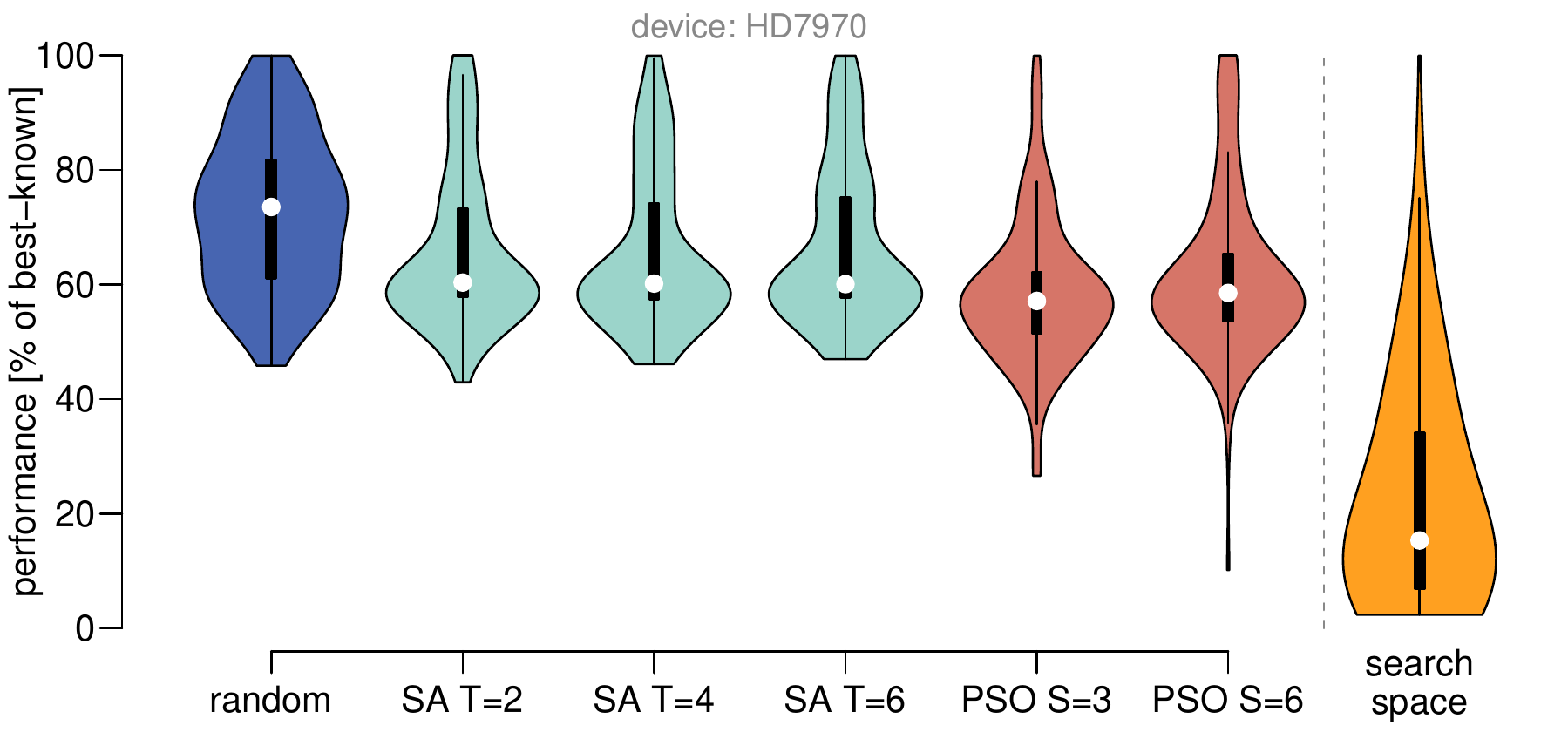}
    \includegraphics[width=0.48\textwidth]{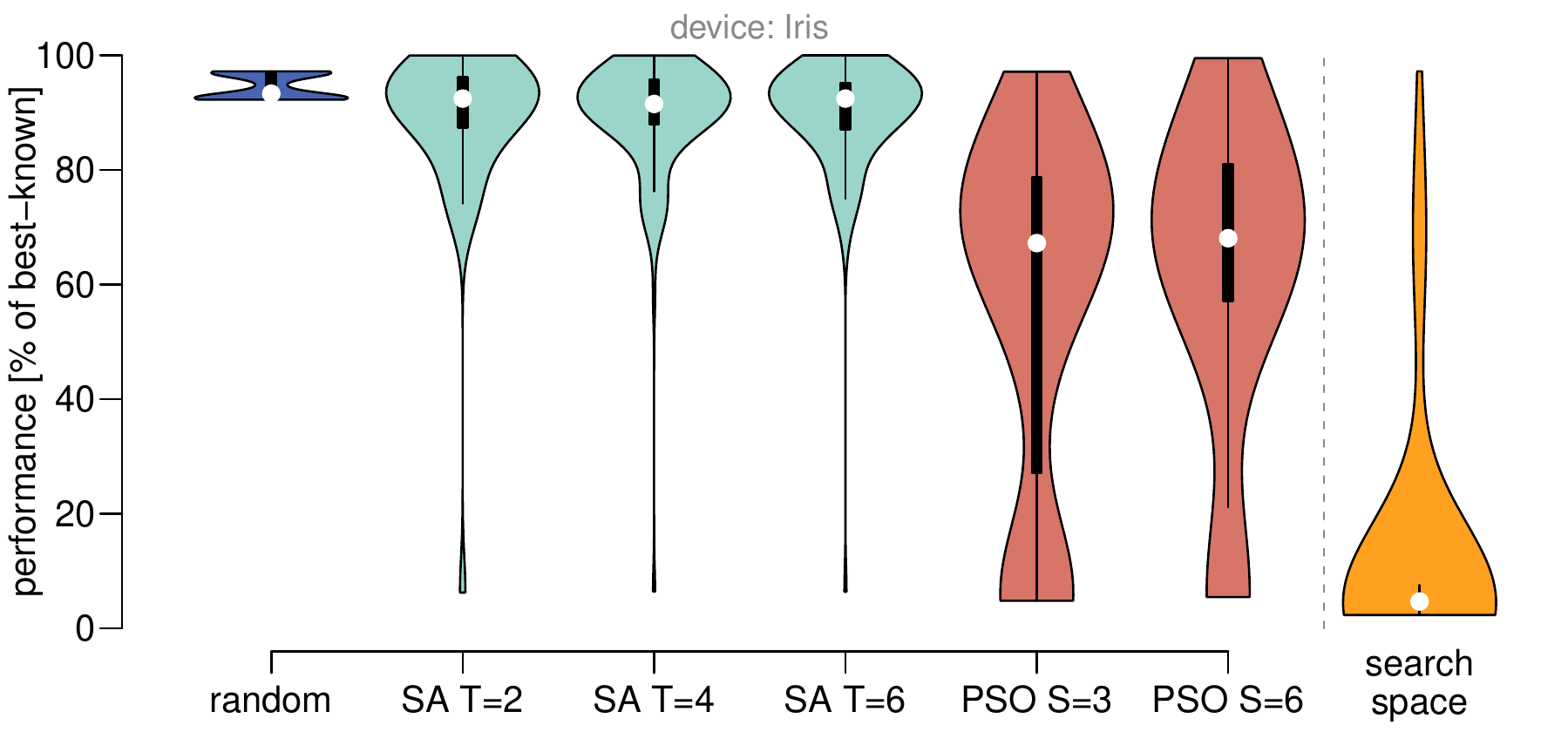}
    \caption{Statistical properties of 128 runs of different search strategies on different devices for 2D convolution. The violin plot shows the standard-deviation (black rectangle), average (white circle), and a rotated density distribution. Note that the search-results displayed represent the best found configurations of each run, not all explored configurations. The right hand side does display the distribution of all possible configurations: the full search-space.}
    \label{fig:conv_violin}
\end{figure*}

Before we evaluate the performance of the best-found results for 2D convolution, we analyse CLTune's search strategies. Within a search-space of 3424 permutations (see table~\ref{tbl:conv} for the values), random-search, simulated annealing and PSO are each configured to explore 107 unique ($1/32^{th}$) configurations. We use an image of 8192 by 4096 and a filter of 7x7 (later also 3x3 and 11x11).

To illustrate the behaviour of the search strategies, we present the search progress of various runs on a single device in Fig.~\ref{fig:conv_trace}. On the left, we see progress traces of 3 runs (in different colours) of the random-search strategy. Each search evaluates 107 configurations and picks the best as its final result, obtaining 56\%, 74\% and 94\% of the best-known performance. The middle plot in Fig.~\ref{fig:conv_trace} holds the results of 3 simulated annealing runs with $T=4$. It shows a general tendency to move towards better configurations, but also shows that simulated annealing can jump to worse configurations when the temperature is still high. The right plot illustrates 3 runs of PSO with a swarm of 3 particles (solid, dashed, dotted), in which each particle visits $107/3$ configurations. Careful observation shows that particles indeed have a tendency to move towards the swarm's best known position.

Because the evaluated search strategies are based on stochastic variables, we cannot draw conclusions from this limited set of experiments. Therefore, we ran each search experiment 128 times on different devices and with different parameters. For clarification: one search experiment explores 107 configurations. The results are shown in Fig.~\ref{fig:conv_violin} as a violin plot: a combination of a boxplot and a rotated kernel density plot. We make several observations based on the results:
\begin{itemize}
    \item Auto-tuning pays off for 2D convolution for all devices: the distribution of the entire search-space (orange rightmost violin) shows that there are only very few configurations with good performance. In fact, setting the parameters to a random configuration yields a performance of less than 20\% on average on each device.
    \item Random search on $1/32^{th}$ of the search-space is already sufficient to achieve an average performance of 77\% on HD7970, 78\% on K40m, 83\% on GTX480, and as high as 92\% on Iris. If that is not sufficient, then simulated annealing or full-search have to be used.
    \item Simulated annealing performs better on average compared to random-search for the K40m and the GTX480. In some cases it shows bad worst-case performance, since the search might get stuck in a local optimum. The algorithm is moderately sensitive to $T$ for these experiments.
    \item PSO performs worse than random-search and is thus not suitable for this search-space. Future work will investigate how its performance changes given a larger swarm size and a longer search (more tested configurations).
\end{itemize}

\subsection{Best-found results}

The tuning parameters and their possible values are shown on the left in table~\ref{tbl:conv}. The right hand side gives the best-found results for 3 different filter sizes and for two devices. These results again demonstrate the usefulness of auto-tuning: the optimal parameters vary across different filter sizes and across different devices, even those from the same vendor.

\begin{table}[!ht]
  \setlength{\tabcolsep}{5pt}
  \centering
  \caption{Best-found parameters for 2D convolution}
  \begin{tabular}{c|c|ccc|ccc|}
                            &                & \multicolumn{6}{c|}{best parameters per filter size} \\
                            & allowed        & \multicolumn{3}{c|}{Tesla K40m} & \multicolumn{3}{c|}{GeForce GTX480} \\
    parameter(s)            & values         & 3x3 & 7x7 & 11x11 & 3x3 & 7x7 & 11x11 \\
    \hline
    $X_{wg}$, $Y_{wg}$      & \{8,16,32,64\} & 32,8 & 32,16 & 32,8    &    64,8 & 32,8 & 32,8 \\
    $X_{wpt}$, $Y_{wpt}$    & \{1,2,4,8\}    & 1,8 & 2,4 & 2,8    &    1,4 & 2,8 & 2,4 \\
    $L\$$                   & \{0,1,2\}      & 0 & 2 & 2    &    0 & 2 & 1 \\
    $VW$                    & \{1,2,4,8\}    & 1 & 2 & 2    &    1 & 2 & 2 \\
    $PAD$                   & \{0,1\}        & 0 & 1 & 1    &    0 & 0 & 0 \\
    $UNR$                   & \{yes,no\}     & yes & yes & yes & yes & yes & yes \\
  \end{tabular}
  \label{tbl:conv}
\end{table}

\begin{figure}[!ht]
    \centering
    \includegraphics[width=0.49\columnwidth]{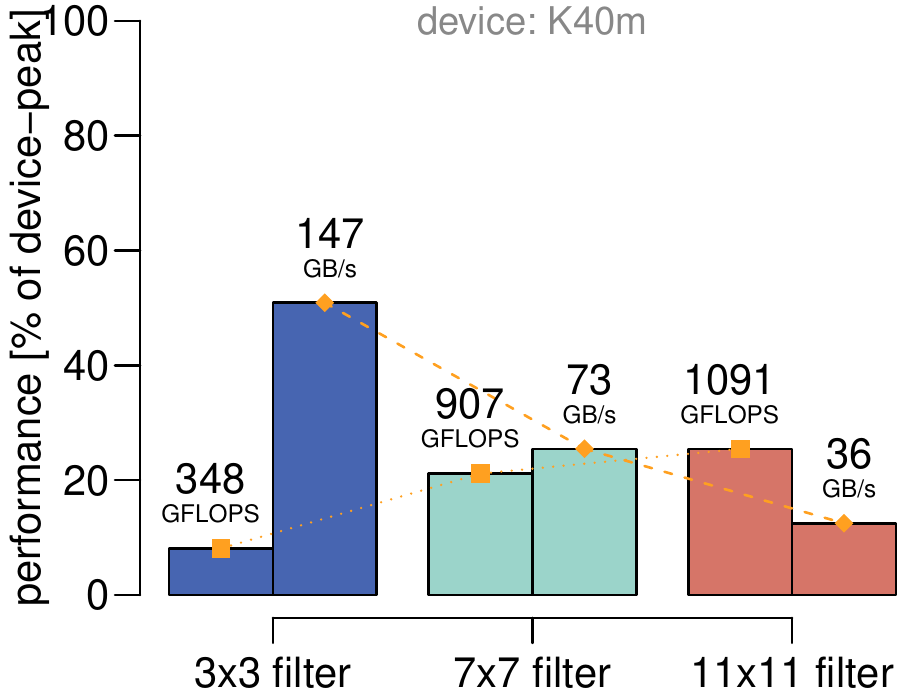}
    \includegraphics[width=0.49\columnwidth]{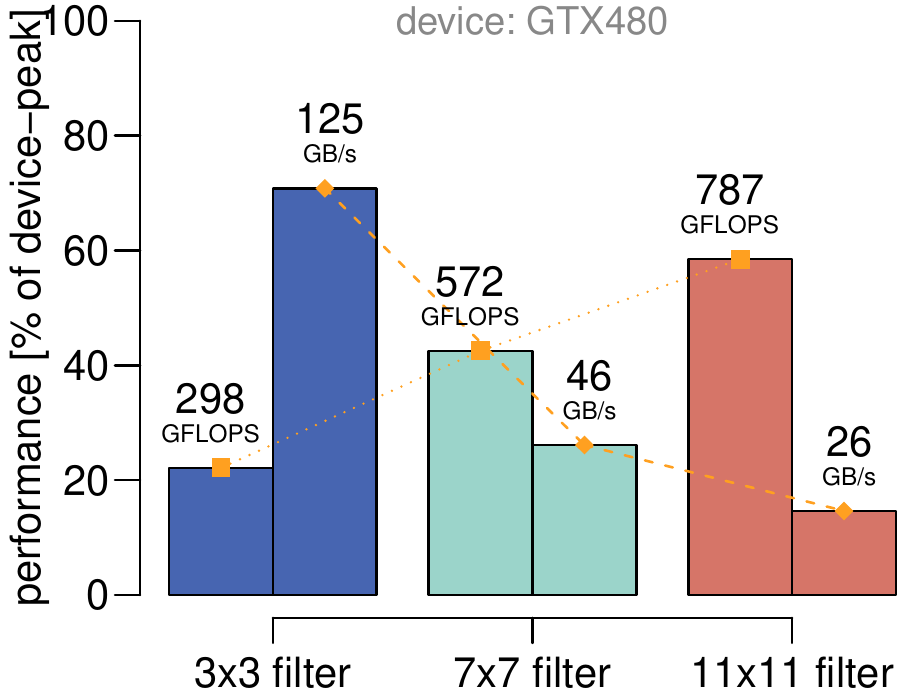}
    \includegraphics[width=0.49\columnwidth]{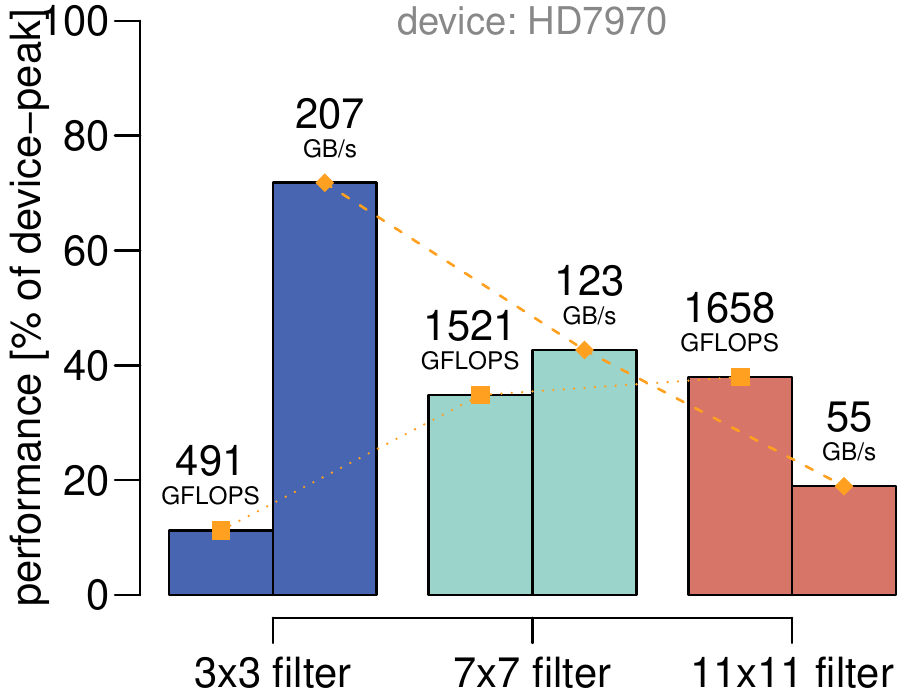}
    \includegraphics[width=0.49\columnwidth]{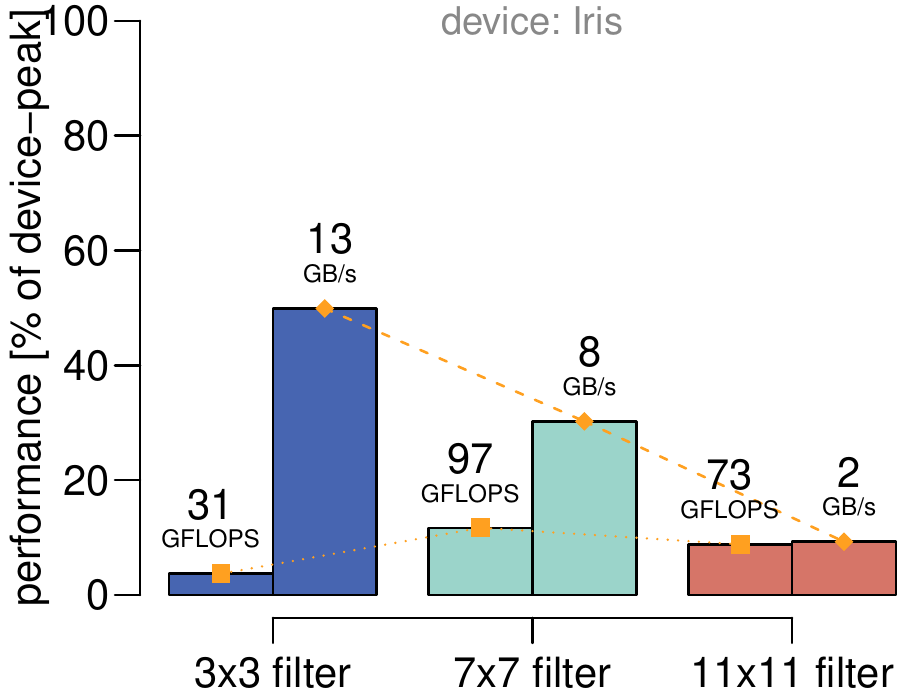}
    \caption{Percentage of the theoretical peak arithmetic throughput (in GFLOPS) and memory bandwidth (in GB/s) for different filter sizes.}
    \label{fig:conv_barplot}
\end{figure}

To demonstrate that our 2D convolution case-study is realistic and useful, we also investigate the absolute performance and compare against related work. Figure~\ref{fig:conv_barplot} first shows the performance relative to the theoretical architectural peak both in terms of GFLOPS and memory bandwidth\footnote{GFLOPS computed as $\frac{(1 + 2 \cdot X_{f} \cdot Y_{f}) \cdot X \cdot Y}{t}$ and bandwidth as $\frac{2 \cdot X \cdot Y}{t}$ with $t$ being the execution time.}. We observe that our implementation follows convolution's mathematical property of becoming more compute intensive as filter sizes are increased. We also note that the GTX480 and HD7970 perform relatively better compared to the K40m due to their better balanced architecture. The Iris GPU is limited by its low memory bandwidth for 3x3 and 7x7. For 11x11, it suffers from a low register count.

To demonstrate the merits of tuning for a specific filter size, we evaluated the best-case parameters for the 3 filter sizes as found in table~\ref{tbl:conv} on the other filter sizes. The results are shown in table~\ref{tbl:filtersizes}, in which relative performance is shown. In the worst case, only 64\% of the performance is achieved: 56\% performance can be gained when running an 11x11 filter with parameters tuned for a 3x3 filter.

\begin{table}[!ht]
  \setlength{\tabcolsep}{8pt}
  \centering
  \caption{Merits of filter size tuning}
  \begin{tabular}{c|ccc|}
    applied to a    & \multicolumn{3}{c|}{best parameters for} \\
    filter of size  & 3x3   & 7x7   & 11x11 \\
    \hline
    3x3       & 100\% &  82\% &  64\% \\
    7x7       &  65\% & 100\% &  83\% \\
    11x11     &  66\% &  75\% & 100\% \\
  \end{tabular}
  \label{tbl:filtersizes}
\end{table}

The state-of-the-art in 2D convolution uses adaptive-tiling, a form of auto-tuning~\cite{Werkhoven2014}. In their work, the authors also experiment on a GTX480 and achieve 326 GFLOPS for 3x3, 550 GFLOPS for 7x7, and 601 GFLOPS for 11x11. As shown on the right hand side of Fig.~\ref{fig:conv_barplot}, we are able to match these numbers for 3x3 and 7x7 and even improve upon them for 11x11. Our improved performance is because they explore a smaller search space: no $Y_{wpt}$, no vector loads and stores, and no extra halo threads. An alternative to 2D convolution is an FFT in frequency space, but as discussed in~\cite{Werkhoven2014}, 2D convolution is significantly faster for filter sizes below 19 by 19 compared to the highly optimized cuFFT library.

\section{Case-study: matrix-multiplication}
\label{sec:gemm}

Our second case-study concerns generalised dense matrix-matrix multiplication (GEMM) as found in the BLAS linear algebra libraries. This case-study is motivated partly by the same reasons as 2D convolution: matrix-multiplication is one of the key components of deep learning and other machine learning algorithms. Furthermore, its wide applicability and FLOP-heavy computation make it a popular target for auto-tuning. Some examples of recent OpenCL auto-tuning work on GEMM are~\cite{Li2009,Matsumoto2012} and the clBLAS library.

In this case-study, we consider the multiplication of two matrices $\mathbf{A}$ and $\mathbf{B}$ according to:
\begin{equation*}
    \mathbf{C} = \alpha \mathbf{A}^{T} \mathbf{B} + \beta \mathbf{C}
\end{equation*}
in which $\alpha$ and $\beta$ are constants and $\mathbf{A}^{T}$ is the transposed version of $\mathbf{A}$. Apart from expecting a transposed input, we also assume that the matrix dimensions are powers of 2 and multiples of the tile sizes (see below). These assumptions can be resolved by a relatively-cheap pre-processing kernel, as is also suggested in~\cite{Matsumoto2012}.

\begin{figure*}[!t]
    \centering
    \includegraphics[width=0.48\textwidth]{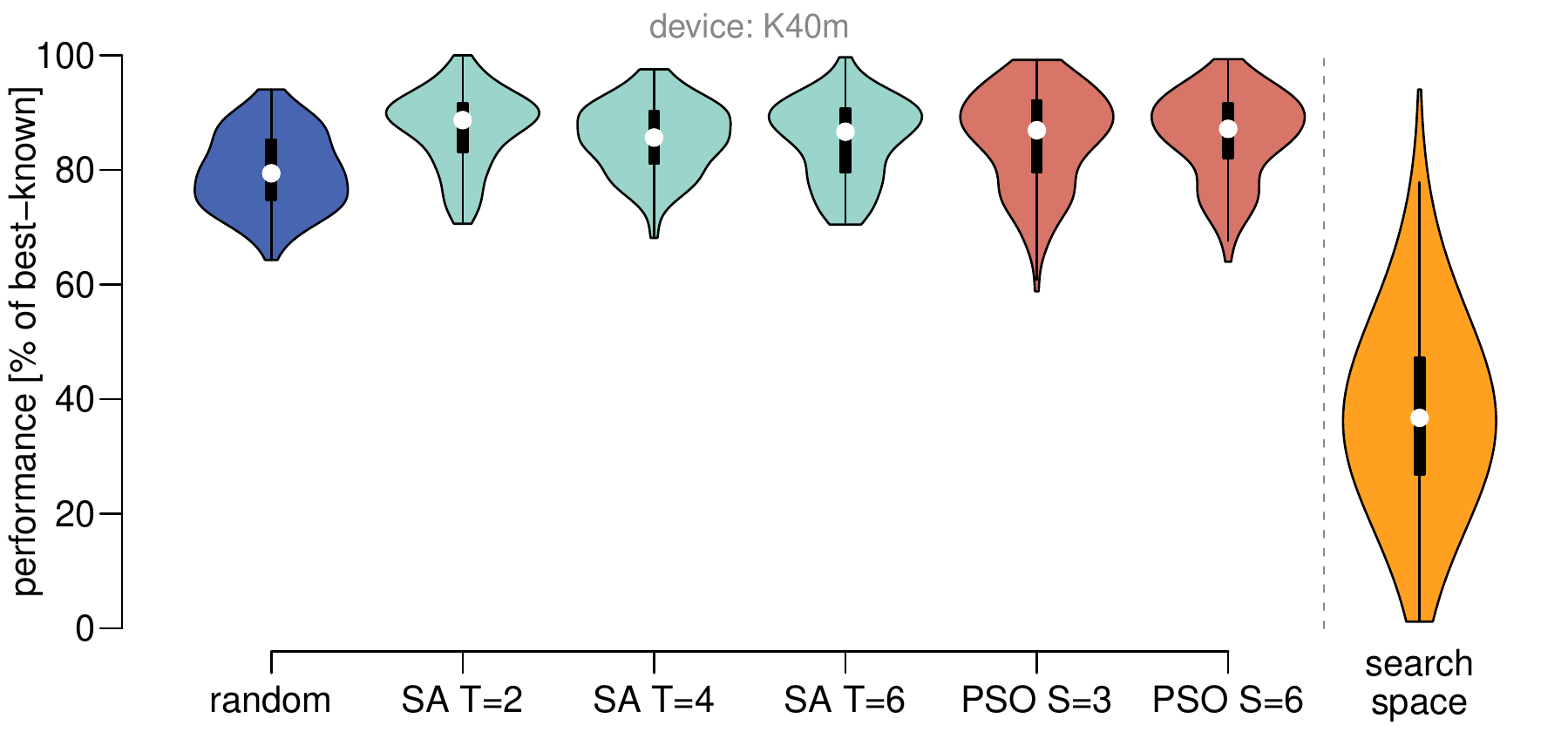}
    \includegraphics[width=0.48\textwidth]{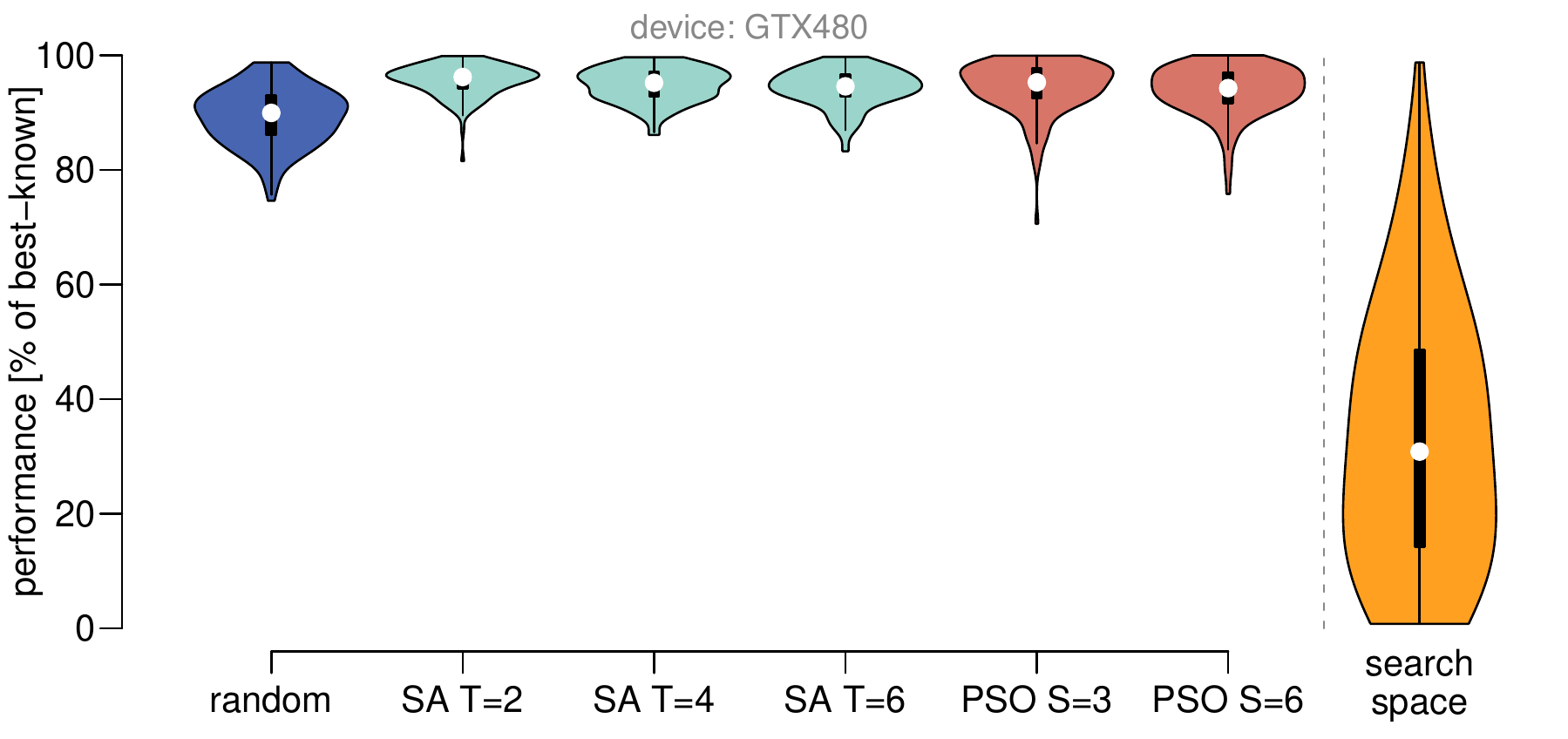}
    \includegraphics[width=0.48\textwidth]{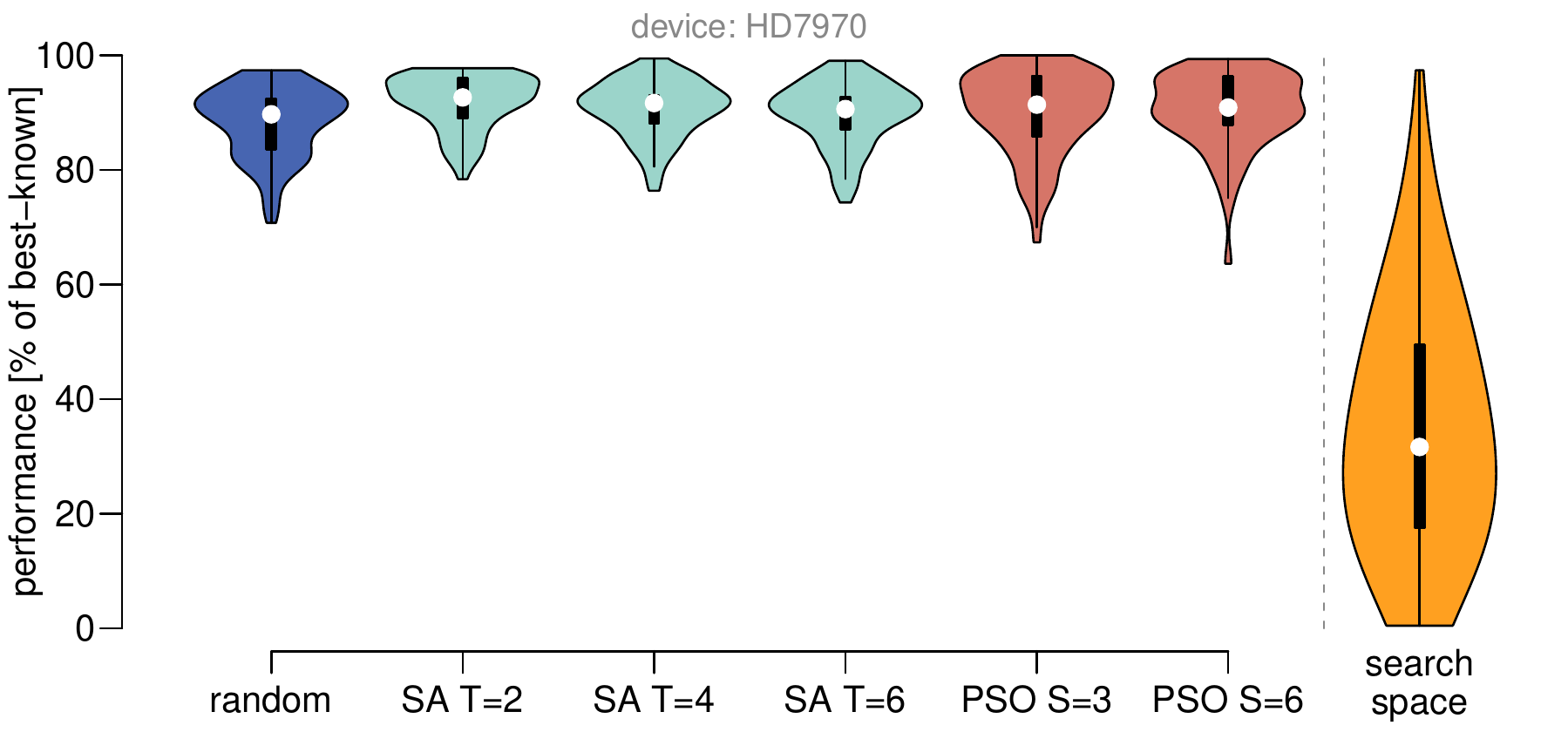}
    \includegraphics[width=0.48\textwidth]{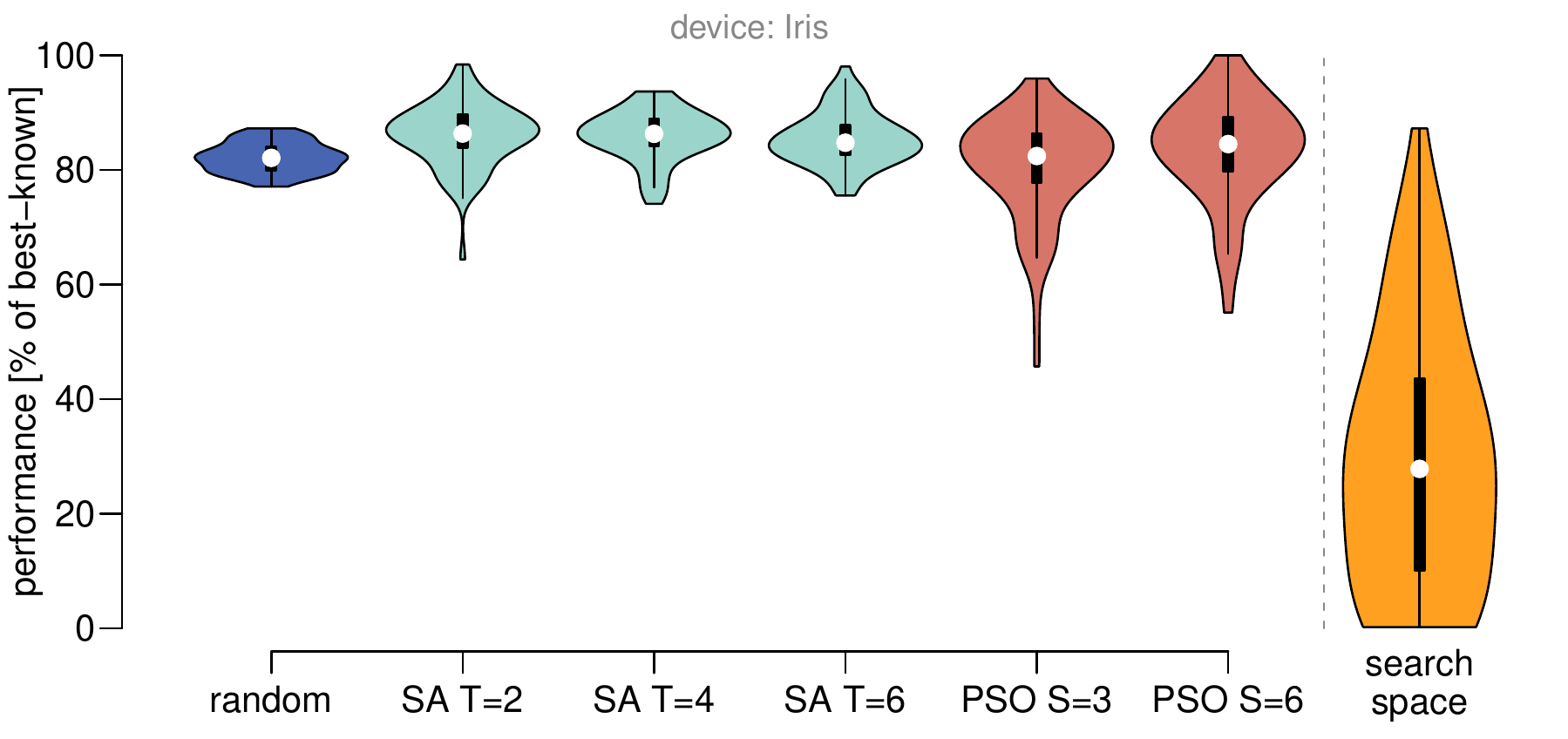}
    \caption{Statistical properties of 128 runs of different search strategies on different devices for matrix-multiplication. See Fig.~\ref{fig:conv_violin} for details about the plots.}
    \label{fig:gemm_violin}
\end{figure*}

\subsection{Tuning parameters}

\begin{figure}[!ht]
    \centering
    \includegraphics[width=0.95\columnwidth]{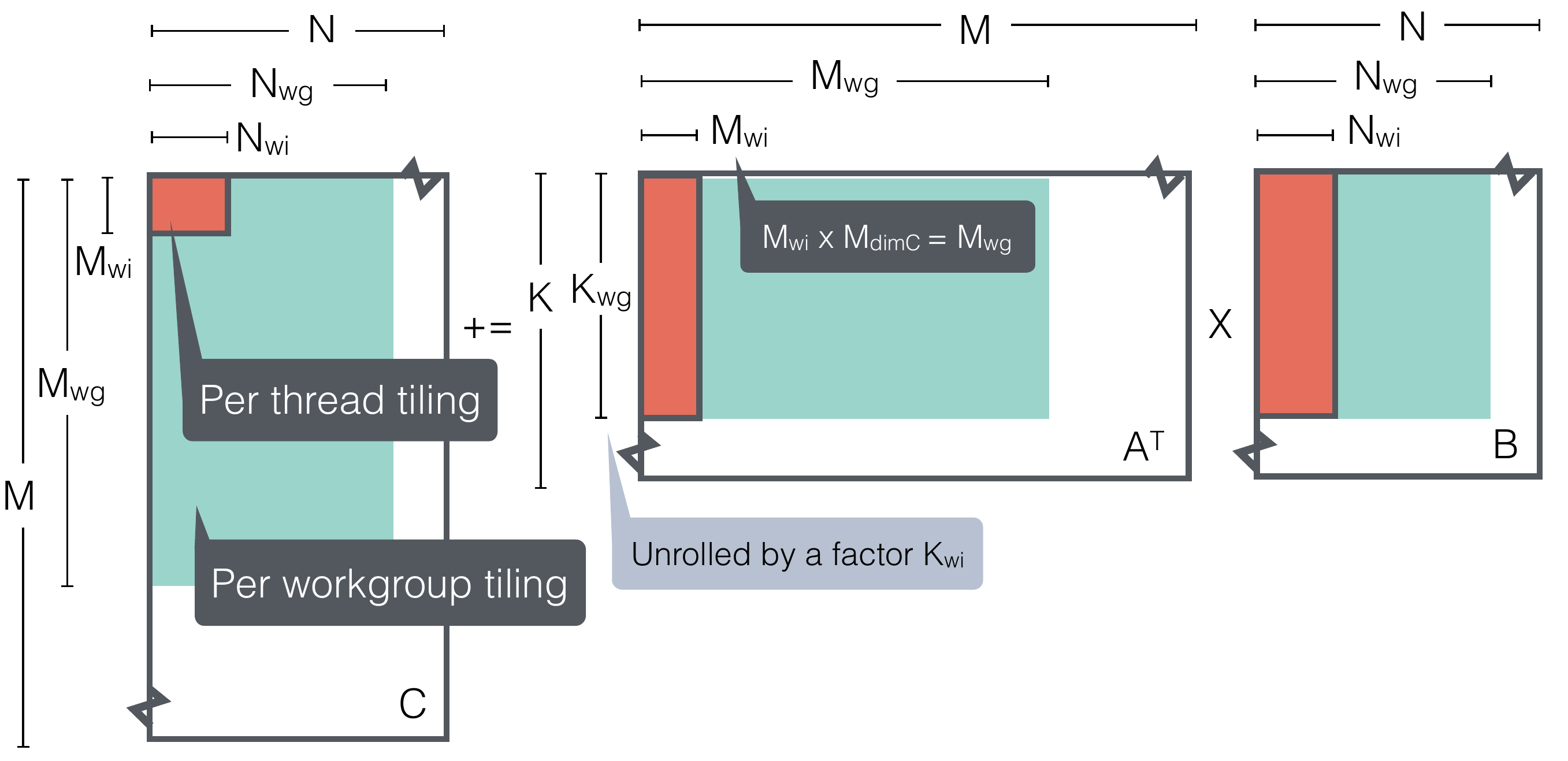}
    \caption{OpenCL matrix-multiplication and some of its tuning parameters.}
    \label{fig:gemm1}
\end{figure}

We implemented a highly tunable parallel version of matrix-multiplication in OpenCL, inspired by~\cite{Matsumoto2012} and the clBLAS library. As far as possible, we use the same parameter names as in~\cite{Matsumoto2012}. The tuning parameters are illustrated in Fig.~\ref{fig:gemm1} and described below:
\begin{enumerate}
    \item 2D tiling is employed at workgroup-level using the parameters $M_{wg}$, $N_{wg}$, $K_{wg}$ corresponding to the $M$, $N$ and $K$ matrix dimensions.
    \item The local workgroup size is tunable in 2 dimensions: $M_{dimC}$ and $N_{dimC}$. Combined with the two corresponding tiling parameters, this also defines the amount of work-per-thread in the $M$ and $N$ dimensions: $M_{wi} = M_{wg}/M_{dimC}$ and $N_{wi} = N_{wg}/N_{dimC}$. Here, $wi$ is an abbreviation for \textit{workitem} (a thread), since coarsening is implemented using 2D register tiling at thread-level.
    \item Caching of the 2D workgroup tile can be controlled per input matrix using $L\$_{A}$ and $L\$_{B}$: manual caching using the local memory is enabled when set to \textit{yes}.
    \item The local memory (when enabled) can be re-shaped according to $M_{dimC} \cdot N_{dimC} = M_{dimA} \cdot K_{dimA} = K_{dimB} \cdot N_{dimB}$. Here, $M_{dimA}$ and $N_{dimB}$ are extra tuning parameters and $K_{dimA}$ and $K_{dimB}$ are calculated according to the above equality.
    \item A stride for accessing off-chip memory within a single thread can be enabled or disabled through $M_{stride}$ (for matrices A and C) and $N_{stride}$ (for matrix B). If enabled, the stride is set to $M_{dimA}$ and $N_{dimB}$ respectively, otherwise it is set to 1 (no stride).
    \item The vector widths for loading and storing can be set using $M_{vec}$ for matrices A and C and $N_{vec}$ for matrix B.
    \item The $K_{wg}$ kernel-loop can be unrolled by a factor $K_{wi}$.
\end{enumerate}

\subsection{Evaluation of the search strategies}

As for 2D convolution, we first evaluate CLTune's search strategies. We use the same devices as listed in table~\ref{tbl:setup} and use the same configurations of the same set of search strategies. For matrix-multiplication, our search-space consists of 241.600 unique configurations: 71x larger than for 2D convolution. The tested values are shown on the left in table~\ref{tbl:gemm}. To keep our search experiments comparable in terms of run-time to 2D convolution, we decide to explore only $1/2048^{th}$ of the search-space: 117 configurations. Our experiments consider squared matrices of 2048 by 2048 ($M=N=K=2048$).

As before, we evaluate the search strategies by running each search 128 times. The violin plots are presented in Fig.~\ref{fig:gemm_violin}. On top of the observations for convolution, we conclude that:
\begin{itemize}
    \item It is easier to find a good performing configuration compared to 2D convolution, judging by the average performance of the search space.
    \item Simulated annealing and PSO perform well on all devices, in all cases outperforming the random-search strategy.
    \item The GTX480 has a better balanced architecture for GEMM compared to the newer K40m, judging by the wider top of the search-space violin (orange, right hand side) and the high average and low standard deviation for the various search strategies.
\end{itemize}

\subsection{Best-found results}

The best-found parameters for single-precision matrix-multiplication are given in table~\ref{tbl:gemm}. The table shows a significant variety in best parameters across the four different devices, again demonstrating the merits of CLTune.

\begin{table}[!ht]
  \setlength{\tabcolsep}{3pt}
  \centering
  \caption{Best-found parameters for matrix-multiplication}
  \begin{tabular}{c|c|cccc|}
                                 & allowed         & \multicolumn{4}{c}{best parameters per device} \\
    parameter(s)                 & values          & K40m & GTX480 & HD7970 & Iris \\
    \hline
    $M_{wg}$, $N_{wg}$, $K_{wg}$ & \{16,32,64,128\}& 128,128,16 & 64,64,32 & 128,128,32 & 64,64,16 \\
    $M_{dimC}$, $N_{dimC}$       & \{8,16,32\}     & 16,16 & 8,16 & 16,16 & 8,8 \\
    $L\$_{A}$, $L\$_{B}$         & \{yes,no\}      & yes, yes & yes, yes & yes, yes & yes, yes \\
    $M_{dimA}$, $N_{dimB}$       & \{8,16,32\}     & 32,16 & 32,32 & 32,32 & 8,16 \\
    $M_{stride}$, $N_{stride}$   & \{yes,no\}      & yes, no & yes, no & no, yes & yes, yes \\
    $M_{vec}$, $N_{vec}$         & \{1,2,4,8\}     & 2,1 & 2,2 & 4,4 & 4,4 \\
    $K_{wi}$                     & \{2,8\}         & 8 & 8 & 2 & 8 \\
  \end{tabular}
  \label{tbl:gemm}
\end{table}

\begin{figure}[!ht]
    \centering
    \includegraphics[width=0.95\columnwidth]{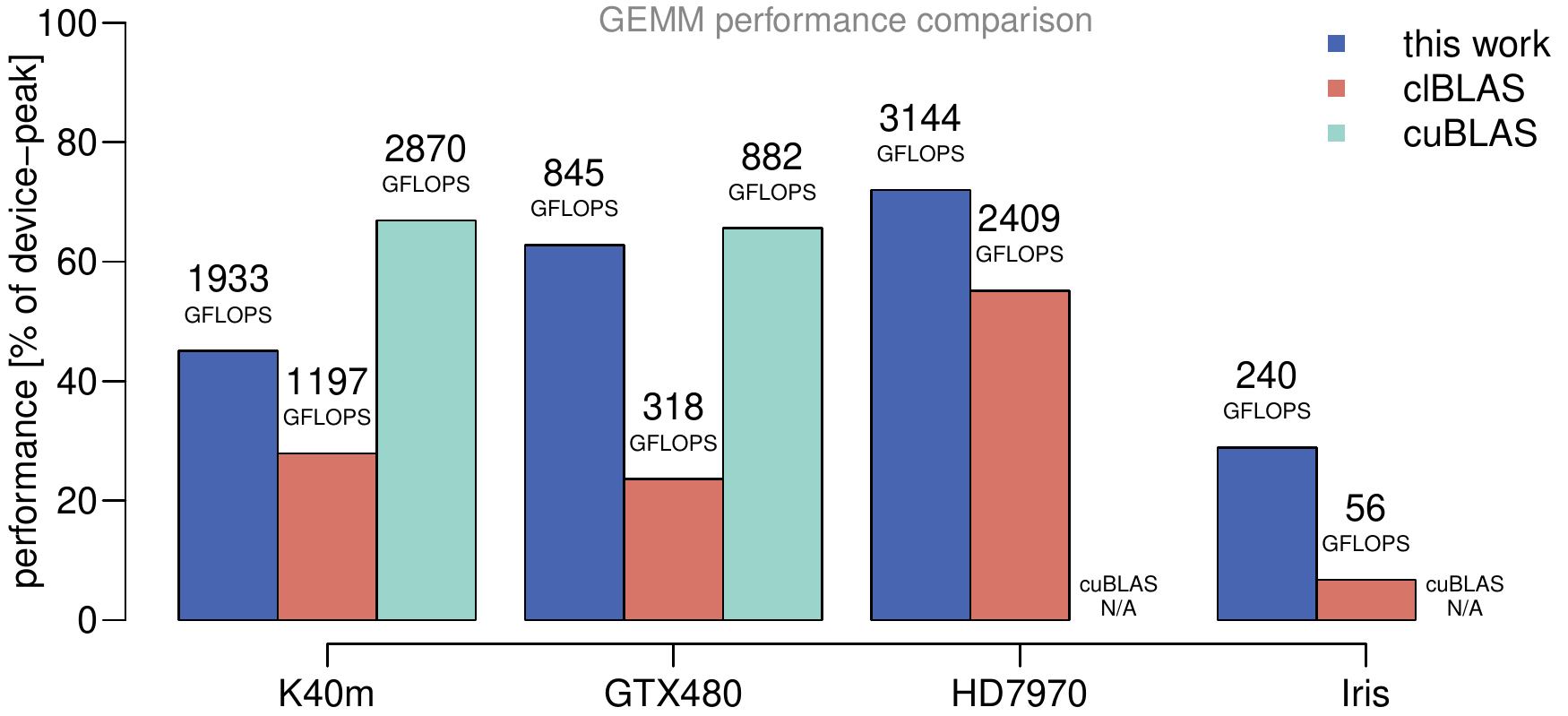}
    \caption{Comparison of our matrix-multiplication against cuBLAS/clBLAS.}
    \label{fig:gemm_barplot}
\end{figure}

The best-found results are compared against the peak theoretical capabilities of the devices and against two libraries: cuBLAS 7.0 for K40m and 5.5 for GTX480, and clBLAS 2.4.0 after running the included tuner. The results are shown in Fig.~\ref{fig:gemm_barplot}, which shows that our matrix-multiplication outperforms the clBLAS library in all cases, including the AMD GPU for which clBLAS was originally developed. We are not able to match cuBLAS on the K40m, as it uses assembly-level optimisations to reduce register pressure and remove register-bank conflicts~\cite{Lai2013}. Furthermore, we are not able to use CUDA's \texttt{ldg} instruction.

To demonstrate the merits of tuning for a specific device, we evaluated all best-case parameters from table~\ref{tbl:gemm} on the K40m GPU. Using the parameters of the GTX480, HD7970 and Iris, we found a performance of respectively 50\%, 61\% and 79\% of the K40m's best results. In other words, up to a factor of 2 can be gained by tuning for a specific device.

Our GEMM implementation is roughly on-par with the auto-tuning work by Matsumoto et al.~\cite{Matsumoto2012,Matsumoto2014}: they reach 2913 GFLOPS on a 18\% lower clocked HD7970. Unfortunately, their OpenCL kernel implementation is not publicly available. Other work improves upon cuBLAS 4.0 by using assembly-level optimisations for NVIDIA GPUs~\cite{Lai2013}, which are now integrated in the tested versions of cuBLAS.

\section{Conclusions}

This work introduced the CLTune auto-tuner for OpenCL kernels. We saw that the generic and open-source CLTune is easy to use and can be used for off-line or on-line tuning. Based on two case-studies and an evaluation on 4 different devices (Tesla K40m, GeForce GTX480, Radeon HD7970, Iris 5100), we conclude that the two advanced search strategies simulated annealing and particle swarm optimisation both have their merits, although their efficacy depends on the problem at hand. We furthermore demonstrated the merits of CLTune by 1) exploring a search space of more than two-hundred thousand configurations, 2) identifying significant differences between best-case parameters across devices, and 3) showing that an OpenCL 2D convolution code can be tuned to input arguments.

With the help of CLTune, we created two OpenCL kernels which are able to match or even improve upon the state-of-the-art: 2D convolution and matrix-multiplication (GEMM). Both have an important advantage over existing work: performance portability and availability on all OpenCL devices. Furthermore, our 2D convolution code is the only tuned implementation available on non-CUDA devices, and our matrix-multiplication is the fastest publicly available source-level implementation for GPUs.


\bibliographystyle{abbrv}
\bibliography{references}

\end{document}